\documentclass{aa}
\usepackage{epsfig,amsmath,amssymb}

\def\mjup{M_{\rm J}}

\def\te{T_{\rm eff}}

\def\simgr{\,\hbox{\hbox{$ > $}\kern -0.8em \lower 1.0ex\hbox{$\sim$}}\,}
\def\simle{\,\hbox{\hbox{$ < $}\kern -0.8em \lower 1.0ex\hbox{$\sim$}}\,}
\def\beq{\begin{equation}}
\def\eeq{\end{equation}}

\begin{document}

%\thesaurus{08.05.03, 08.22.1}

\title{The effect of evaporation on the evolution of close-in giant planets}
 \author{I. Baraffe\inst{1}, F. Selsis\inst{2}, G. Chabrier\inst{1}, T. S. Barman\inst{3},
F. Allard\inst{1}, P.H. Hauschildt\inst{4} and H. Lammer\inst{5}
}

\offprints{I. Baraffe}

\institute{C.R.A.L (UMR 5574 CNRS),
 Ecole Normale Sup\'erieure, 69364 Lyon
Cedex 07, France (ibaraffe, chabrier, fallard@ens-lyon.fr)
\and 
Centro de Astrobiolog\'{\i}a (INTA-CSIC), Ctra. de Ajalvir km 4, 28850 Torrej\'on de Ardoz, Madrid, Spain (selsis@obs.u-bordeaux1.fr)
\and
Department of Physics, Wichita State University, Wichita, KS
67260-0032 (travis.barman@wichita.edu)
\and
Hamburger Sternwarte, Gojenbergsweg 112,
21029 Hamburg, Germany (yeti@hs.uni-hamburg.de)
%Center for Simulational Physics, University of Georgia
%Athens, GA 30602-2451 (travis, yeti@hobbes.physast.uga.edu)}
\and
Space Research Institute, Austrian Academy of Sciences, Schmieldstrasse 6, A-8042 Graz, Austria (helmut.lammer@oeaw.ac.at)}

\date{Received /Accepted}

\titlerunning{}
\authorrunning{Baraffe et al.}
\abstract{We include the effect of evaporation in our evolutionary calculations of close-in giant planets, based on a recent
model for thermal evaporation taking into account the XUV flux of the parent star (Lammer et al. 2003). 
Our analysis leads to the existence of a critical mass for a given orbital distance
 $m_{\rm crit}(a)$
below which the evaporation timescale becomes shorter than the thermal timescale of
the planet. For planets with
initial masses below $m_{\rm crit}$, evaporation leads to a
rapid expansion of the outer layers and of the total planetary radius, speeding up
the evaporation process. Consequently, the planet does not survive as 
long as estimated by a simple application of mass loss rates without 
following consistently its evolution. We find out that  the
transit planet HD 209458b might be in such a dramatic phase, 
although with an extremely small probability. As a consequence, we predict
that, after a certain time, only planets above a value $m_{\rm crit}(a)$ should be present at an
orbital distance $a$ of a star. For planets with initial masses above $m_{\rm crit}$,
evaporation does not affect the evolution of the radius with time.  
\keywords{ planetary systems --- stars: individual (HD209458, OGLE-TR-56)} 
}

\maketitle

\section{Introduction}

The increasing number of discovered giant planets orbiting at $\simle$ 0.1 AU from
their parent star raises fundamental questions about their formation
and migration process and about the influence of the parent star through irradiation or 
tidal effects. The recent discovery of an extended atmosphere for the transiting exoplanet HD209458b (Vidal-Madjar et al. 2003) highlights the occurence of atmospheric evaporation
for these close-in planets.
%Among the various effects resulting from the central star incident flux, escape of the hydrogen rich atmosphere could bear important consequences
%on the structural properties of the planet. 
Whether such evaporation due to heating from the incident stellar flux leads to major mass loss during the planet lifetime, and whether this process affects significantly the structure of
 the planet and thus its $m$-$R$ relationship is an open question, which is of prime importance for our understanding of planetary system formation. 
%Indeed, the currently observed distribution of planet masses as a function of their orbital separation is unlikely to reflect the initial mass distribution and may thus lead to erroneous formation scenarios.
% for the less massive planets. 
%Given the unexpectedly large radius of HD209458b for its observed mass (Guillot \& Showman 2002, Baraffe et al. 2003, Burrows et al. 2003), one thus wonders whether evaporation . 
New evaluations of atmospheric thermal evaporation rates by Lammer et al. (2003, L03), based on 
exospheric heating by stellar XUV radiation, yield significantly larger rates
than the previous estimates assuming Jeans escape at the effective temperature of the planet.
%the exospheric temperatures determined by the XUV radiation of
%the parent star, yield significantly larger rates than previous estimates, based on the much cooler planet effective temperature. 
The first attempt of L03 to model such a complex process 
%and based on this new idea
%is very promising since 
yields an escape rate in good agreement with the observational 
determinations of 
Vidal-Madjar et al. (2003) for HD209458b, providing encouraging support for further exploration. Moreover, since stellar XUV fluxes vary significantly with time and can be
order of magnitudes larger at very young ages, these evaporation rates are much larger at the planet early evolutionary
stages.
%Given the success of their model applied to HD209458b, 
 L03 thus suggest that mass loss could
be an important event in the life of close-in exoplanets, 
contrarily to what was previously thought. 
It is the purpose of this letter to explore this issue
%analyse the effects of evaporation on close-in exoplanets
by taking into account consistently the thermal escape rates of
L03 along the evolution of strongly irradiated planets (Baraffe et al. 2003, hereafter B03).
%thus including evaporation effects in recently developed consistent irradiated evolutionary calculations (Baraffe et al. 2003, hereafter B03). 
%Such models take
%into account irradiation effects from the incident stellar flux on the atmospheric 
%and inner structures of the planet. Present calculations thus
%include consistently for the first time in evolutionary calculations irradiation and 
%due to the entire flux distribution of 
%the parent star, including the high energy part of the spectrum.
Since an important issue of this analysis is to determine whether
evaporation affects significantly the inner structure and thus the $m$-$R$ relationship of
exoplanets, we
focus on the case of presently detected transits, namely HD209458b, with
 $a=0.046$AU (Charbonneau et al. 2000) and OGLE-TR-56b, with $a=0.023$AU
 (Konacki et al. 2003).
%Such observations provide unique tests of the mass-radius relationship for
%irradiated exoplanets and thus of our predictions of evaporation effects. 
%Extension to other cases (different parent star spectral types and orbital separations) is under progress.

\section{Description of the models}

%\subsection{Models for irradiated planets}

The evolutionary calculations are based on the consistent coupling between the
irradiated atmospheric and interior structures as described in B03, and in
Barman et al. (2001) for the atmosphere model calculations.
Such a consistent treatment of the irradiated
atmospheric structure and the internal, partially radiative structure successfully reproduces the  observed parameters of the transit planet OGLE-TR-56b (Chabrier et al. 2004).
%, although the large radius of HD209485b, if not due to tidal heating of an undetected companion, requires a more complicated physics (see Chabrier et al. 2004 for discussion).

%\subsection{Thermal evaporation rates}

Details of the model used to derive thermal evaporation rates can be found in L03. The basic idea relies on the fact that
%without cooling due to the expansion and the loss of the upper atmospheric layers, 
the energy deposition by stellar XUV leads to exospheric temperatures higher than the blow-off temperature for H. Therefore, the classical Jeans' description of thermal escape
must be replaced by a hydrodynamic modeling of  the expansion and mass loss. 
The energy-limited atmospheric mass loss rate $\dot M$ can be 
written:

\begin{equation}
\dot M = 3 \beta^3 F_\star / ( G \rho),
\end{equation}

\noindent where $\beta$ is the ratio between the expansion radius $R_1$, where the bulk of the XUV radiation is absorbed,  and the planetary radius $R_{\rm P}$ and $\rho = (3 m)/( 4 \pi R_{\rm P} ^3)$ is the mean planet density. 
L03 estimate $\beta$ by applying the hydrodynamic model of Watson et al. (1981).
 The term $F_\star$
%(in W.${\rm m}^{-2}$) 
 is the stellar flux, averaged over the whole planet surface, taking into account both the contribution in the 1-1000~\AA{} wavelength interval and the Lyman-$\alpha$ flux, 
so that the total contribution $F_\star$ for an orbital separation
$a$ (in AU) is $F_\star = (F_{\rm XUV} + F_\alpha)/a^2$.
%$G$ is the gravitational constant and  
% (in kg.m$^{-3}$). 
Although $\beta$ is expected to vary with  $F_\star$ and $a$,
we fix it at the maximum value $\beta$ = 3 found in L03,
which is in good agreement with the observed expanded exosphere of HD209458b (Vidal-Madjar et al. 2003).
Estimates of XUV and Lyman-$\alpha$ fluxes and their time dependence are based on the 
current solar value and a collection of data for solar type stars covering an age
from 100 Myr to 8 Gyr.
As in L03, we adopt for the XUV 
and  Lyman-$\alpha$ contributions:

\begin{equation}
\begin{cases}
F_{\rm XUV}(t)  = 6.13 \, t^{-1.19} f_{\rm XUV} & \text{if $t \geqslant 0.1$ Gyr,} \\
F_{\rm XUV}(t)  = F_{\rm XUV}(0.1) & \text{if $t<0.1$ Gyr} 
\end{cases}
\end{equation}

\begin{equation}
\begin{cases}
F_\alpha(t) = 3.17 \, t^{-0.75} f_\alpha & \text{if $t \geqslant 0.1$ Gyr,} \\
F_\alpha(t)  = F_\alpha(0.1)  & \text{if $t<0.1$ Gyr} 
\end{cases}
\end{equation}

\noindent where $t$ is the age in Gyr, and $f_{\rm XUV} \, = \, 8.5 \, 10^{-4}$ W.m$^{-2}$
%$f_{\rm XUV} \, = \, 3.4 10^{-3}$/4 W.m$^{-2}$
and $f_\alpha$ = 1.42 10$^{-3}$ W.m$^{-2}$ 
%$f_\alpha$ = 5.7 10$^{-3}$/4 W.m$^{-2}$
are respectively the Sun XUV and
Lyman-$\alpha$ surface-averaged fluxes at 1AU (Woods \& Rottman 2002).
%(Floyd et al. 1999, Lean et al. 2003, Bayley et al. 1999, Woods et al., 2002). 
%These values are taken
%from Floyd et al. (1999) for Lyman-$\alpha$, Lean et al. (2003) for the 500-1000~\AA{} range, Bayley et al. (1999) for the 350-500~\AA{} range and from the recent Solar EUV Experiment (Woods et al., 2002) for  the soft X-rays ($<350$~\AA{}).
%The division by a factor 4 in both expressions stems from the  fact the planetary
%disk intercepts $\pi R^2$ of the solar flux which is then averaged over the total
%planetary surface $4 \pi R^2$. 
Equations (2) and (3)  recover the solar values for $t$=4.5 Gyr. 
%The time dependence of Eq. (2) and (3) are only valid
%for ages $>$ 0.1 Gyr. Since extrapolating such relations at young ages is extremely
%hazardous, we rather fix  $F_{\rm XUV}$ and $F_\alpha$ at their values
%at 0.1 Gyr for $t < 0.1$, keeping in mind that such choice may underestimate the
%stellar contribution at early times. (FRANCK, C'EST VRAI CE QUE JE DIS? TU AVAIS L'AIR DE DIRE PLUTOT QUE LE FLUX ETAIT CONSTANT ET MAXIMAL entre t=0 et 100 MYR, JUSTIFICATION?)
%The total contribution $F_\star$ for any orbital separation
%$a$ (in AU) is thus:
%\begin{equation}
%F_\star = (F_{\rm XUV} + F_\alpha)/a^2
%\end{equation}

\section{Evolution of evaporating planets}

We have implemented the evaporation rates given by
Equ. (1) in our evolutionary code for irradiated planets, focusing on the two aforementioned orbital separations, $a$=0.023 AU and $a$ = 0.046 AU. 
%In all cases, we take into account consistently
%the effect of irradiation on the thermal structure of the planet atmosphere.
We investigate planets with initial masses 0.5 $\mjup$ to 5 $\mjup$, orbiting a solar type star ($T_{eff_\star}=$ 6000 K). 
For this mass range, evaporation rates vary from $\sim$ 10$^{-8} \mjup$/yr at young
ages to $\sim$ 10$^{-12} \mjup$/yr for $t > $ 5 Gyr.

\subsection{Response to mass loss: the case of a 1 $\mjup$ planet}

\begin{figure}
\psfig{file=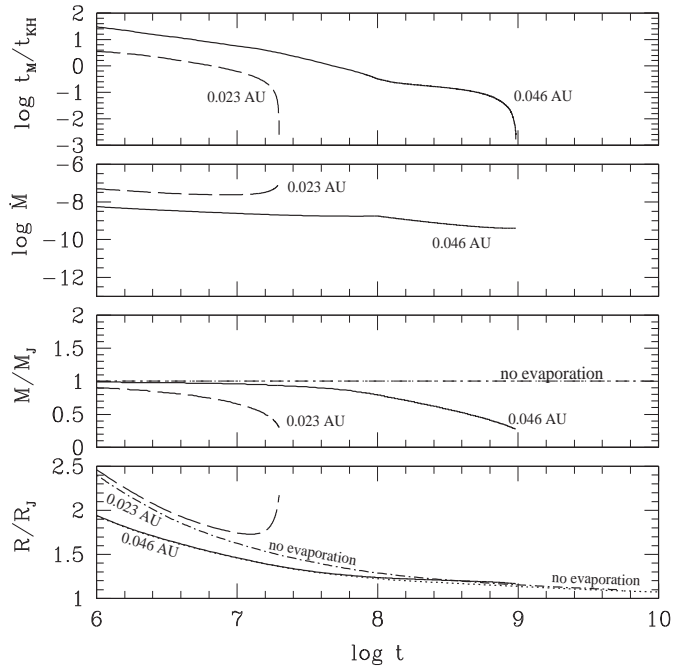,height=88mm,width=88mm} 
\caption{Evolution of an evaporating planet with initial mass 1 $\mjup$ at different
orbital separations. From lower to upper panel: the planet radius (in units
of Jupiter radius $R_{\rm J}$ = 7.1492 10$^9$ cm), mass, evaporation rate $\dot M$ in $\mjup$/yr and the ratio of the evaporation rate to thermal timescale 
$\log \,  (t_{\rm \dot M} / t_{\rm KH} )$. 
The solid (with evaporation) and dotted  (no evaporation, lower panels) curves correspond to $a$ = 0.046 AU. The long-dashed (with evaporation) and dash-dotted  (no evaporation, lower panels) curves correspond to $a=0.023$ AU. 
%The dashed curve in the upper panel corresponds to the case $a$ = 0.046 AU for which evaporation
%started only at $t$ = 20 Myr (see text). 
}
\label{fig1}
\end{figure}

Figure \ref{fig1} shows the evolution of an evaporating planet with initial mass 1$\mjup$. 
A key quantity to understand the response of an object to mass loss
is the ratio of the 
mass loss rate timescale $ t_{\rm \dot M} = m/ \dot M$ to the thermal timescale, characterised by the Kelvin-Helmotz timescale $t_{\rm KH} \sim 2 G m^2/(RL)$.
Such a ratio is displayed in the upper panel of Fig. \ref{fig1}. 
As long as $t_{\rm \dot M}/t_{\rm KH} \, > 1$,
the evolution of the planet radius $R(t)$ is barely affected by evaporation, as seen by the comparison between the non-evaporating and evaporating curves in the lower panel of Fig. \ref{fig1} (see also Fig. \ref{fig3}). 
However, when $t_{\rm \dot M}$ becomes shorter than $t_{\rm KH}$, the evolution
 of the planet changes drastically:  its mass decreases faster than its contracting radius.
%, the decrease of which stems from the contraction of the planet with time. 
Consequently, $m / R_{\rm P}$ 
decreases increasingly rapidly with time, or conversely the planet mean density increases
more and more slowly, compared to a case with slower evaporation. The
evaporation rate thus decreases more slowly with time since $\dot M \propto 1/\rho \propto R_{\rm P}^3/m$ (Eq. (1)).
%due to both the weakening of the XUV flux and the incr%ease of the planet mean density, 
Consequently, the ratio $t_{\rm \dot M} /t_{\rm KH}$ keeps decreasing.
When $t_{\rm \dot M} \simle  t_{\rm KH}/10$, the response of the planet becomes
rather violent, as indicated by the sudden increase of the radius 
at  $t \sim 20$ Myr  for the case $a$ = 0.026 AU (Fig. \ref{fig1},  long-dashed curve). 
This stems from the fact that the outer layers
start to expand significantly and the radius, instead of decreasing slowly or remaining constant,
increases rapidly, yielding a sudden increase of the evaporation rate $\propto R^3_{\rm P}$, producing in turn further
expansion of the outer layers. The situation turns into
kind of a runaway behaviour and points to a catastrophic fate for the planet, at least concerning its
hydrogen-rich  envelope.

The expansion of the outer layers, yielding eventually such a violent reaction, can be understood in terms of entropy balance, in analogy
with mass loss of low mass secondaries in compact  binary systems 
%Stehle et al. 1996; 
(Ritter 1996).
%Such analysis (Stehle et al. 1996; Ritter 1996) show that a fully convective low  mass star undergoing
%rapid mass loss expands compared to its thermal equilibrium configuration.
%Similar arguments based on entropy profiles can be used,
%although the situation for planets is different
%since they are continuoulsy contracting on a thermal timescale, {\it i.e} there is no thermal equilibrium configuration, and they are irradiated. 
The entropy profile of an irradiated planet
of mass $m$ is constant throughout most of the structure.
The outer layers, however,
are radiative, and are characterized by a nearly isothermal, high entropy profile
(Guillot et al. 1996; Barman et al. 2001; B03; Chabrier
et al. 2004). The  mass enclosed in the radiative layers is typically $\sim$ 10$^{-5}$
$\mjup$. With an evaporation rate $\dot M \sim 10^{-8} \mjup$/yr, it takes 
$\Delta t \sim 10^3$ yr
to evaporate all these layers. The upper convective zones of entropy $S_{\rm conv}$ 
are then exposed to irradiation of the parent star, and become radiative with a significantly higher entropy $S_{\rm rad}$. 
In terms of 
gravothermal energy rate defined as:
\begin{equation}
\epsilon_{\rm grav} = - T {\Delta S \over \Delta t} = -T {(S_{\rm rad} - S_{\rm conv}) \over \Delta t},
\end{equation}
\noindent $\epsilon_{\rm grav}$ is thus negative. Gravothermal energy is thus
released in the outer layers, and converted mainly into expansion work. At some time, this
expansion work becomes significant compared with the (positive) gravothermal energy rate
due to the planet contraction and cooling. This occurs
%, which becomes relatively small since it
when $t_{\rm KH}$  $\gg$ $t_{\rm \dot M}$, yielding
%the total radius increases. This thus leads to 
a runaway situation where evaporation is speeding up and leading eventually to a catastrophic event if other processes
do not regulate the increase of $R$ and $\dot M$. 
When a planet reaches such a critical regime, evaporation is speeding up drastically, at a much higher rate than expected from standard scenarios.

%Following the evolution of the planet in this rather unstable phase is numerically difficult and requires very small numerical timesteps. Given the uncertainties of the present evaporation rates (see discussion in \S 4), exploring the consequences of this reaction by detailed numerical calculations is out the scope of the present letter.
As seen in Figure \ref{fig1}, $ t_{\rm \dot M} <t_{\rm KH}$ after 10 Myr for $a$ = 0.023 AU and after 50 Myr for $a$ = 0.046, respectively.
The $a$ = 0.046 sequence shown in Fig. \ref{fig1} stops before showing a strong increase of  $R$ simply because
 it reaches the lower $\te$-limit of our irradiated atmosphere grid, i.e. $\te = 50$ K. 
%We expect, however, a behavior and a fate similar to the planet located at $a$ = 0.023 AU. 
The planet
at $a$ = 0.046 AU survives $\sim$ 1 Gyr and the one at $a$ = 0.023 AU only 20 Myr. Using the same evaporation rates without considering the effects on the evolution yields a time for
complete evaporation of a 1 $\mjup$ planet of $\sim$ 2 Gyr  at $a$ = 0.046 and  $\sim$ 40 Myr
at $a$ = 0.023 AU. A consistent treatment of evaporation and evolution, yielding the aforedescribed evaporation speeding up process, thus decreases appreciably, by a factor of
$\sim 2$, the timescale of a planet for complete evaporation.

\subsection{A critical mass for speeding up evaporation}

\begin{figure}
\psfig{file=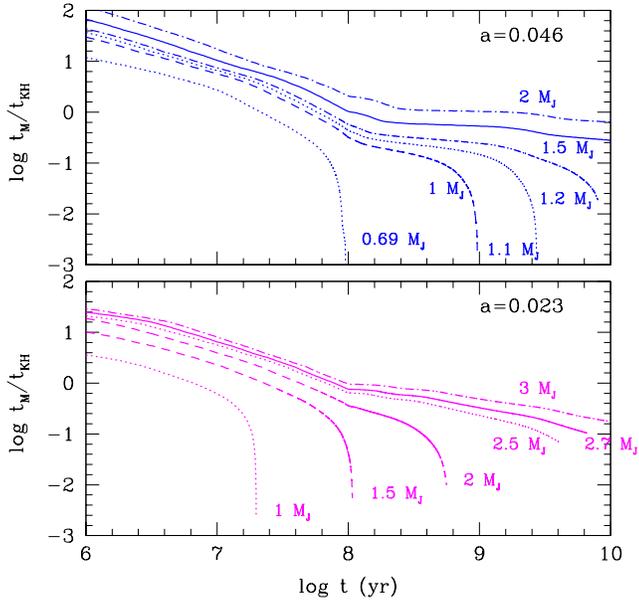,height=88mm,width=88mm} 
\caption{Ratio of the mass loss timescale to thermal timescale 
$t_{\rm \dot M} / t_{\rm KH}$ as a function of time (in yr) for planets with different initial masses, as indicated
on the panels. Upper panel corresponds to orbital separation $a$ = 0.046 AU and
lower panel to $a$ = 0.023 AU. In both panels, the solid curve indicates the critical
mass below which $t_{\rm \dot M} / t_{\rm KH} < 1/10$ in less than 5 Gyr (see text).
}
\label{fig2}
\end{figure}

\begin{figure}
\psfig{file=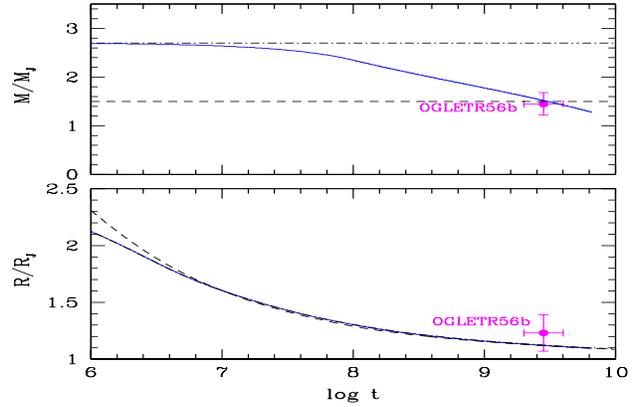,height=60mm,width=88mm} 
\caption{Effect of evaporation on the radius and mass as a function of time
(in yr) for planets at $a$ = 0.023 AU from 
their parent star. The solid curve is an evaporating planet with initial mass 2.7 $\mjup$, which
reaches 1.5 $\mjup$ in 3 Gyr,  reproducing the properties of  OGLE-TR-56b (Torres et al. 2003).
Comparison is made with non evaporating evolutionary sequences with mass 2.7 $\mjup$ ( dash-dot)
and 1.5 $\mjup$ (dash). Note that the solid and dash-dot curves are undistinguishable  in the lower panel.
}
\label{fig3}
\end{figure}

%The results of the previous section show that for a given orbital separation, there is an initial planetary mass below
%which evaporation becomes faster than the thermal evolution. When 
%$t_{\rm \dot M} / t_{\rm KH}\simle 0.1$, further evaporation
%leads to a runaway which speeds up the process. This phenomenon may lead to a catastrophic event for the hydrogen-rich envelope and shorten appreciably the lifetime of the planet.
Following the results of the previous section, we define a critical mass $m_{\rm crit}$ below which $t_{\rm \dot M} = t_{\rm KH}/10$
%, a condition for evaporation to proceed faster than thermal evolution, leading to a runaway evaporation process,
is reached in $<$5 Gyr. A planet with initial mass $m < m_{\rm crit}$ orbiting a Sun-type star
will
thus evaporate entirely (or leave at least a rocky core)  in less than 5 Gyr. 
Based on the present evaporation rates, we find that  $m_{\rm crit} \sim 1.5 M_J$
for $a$ = 0.046 AU and $m_{\rm crit} \sim 2.7 M_J$
for $a$ = 0.023 AU, as illustrated in Figure \ref{fig2}.
Note that in our calculations, evaporation starts at the very beginning of our evolutionary
sequences. This is questionable, since heating from the parent star is likely
to be smaller at the very early stages of the planet evolution,
because either of the presence of a protoplanetary disk or the fact that the planet may form
at a larger orbital distance and migrate inwards.
Both processes have timescales obviously linked to the disk lifetime, with a reasonable upper
limit of $\sim$ 10 Myr (Strom et al. 1993, Armitage et al. 2003). On such short timescales,
evaporation effects on the evolution of the planet are unsignificant in the presently considered mass range.
Indeed, we checked that starting evaporation at an age $t \sim 10$ Myr does not change
the fate of the planet and thus $m_{\rm crit}$.
A prediction of the present calculations is thus the absence of exoplanets below the aforementioned critical masses
and orbital distances after 5 Gyr.

\subsection{Evolution of the radius}

For planets with initial mass $m \simgr m_{\rm crit}$, the evolution of the radius, $R(t)$, is  barely distinguishable from non-evaporating cases. 
This is illustrated in Figure \ref{fig3} which portrays the evolution of a 2.7 $\mjup$ planet
located at $a$ = 0.023 AU from its parent star. With the present evaporation rates, the evaporating sequence
(solid line) reaches a mass 1.5 $\mjup$  and a radius 1.12 $R_{\rm J}$ after 3 Gyr, in agreement
with the observed properties of OGLE-TR-56B (Torres et al. 2003). The dashed curve
shows the evolution of a non-evaporating planet of mass 1.5 $\mjup$, which
 reaches a similar radius 1.11  $R_{\rm J}$ at $t$ =  3 Gyr. At such an age, the evaporation rate is $\sim$
5.5 10$^{12}$ g s$^{-1}$, orders of magnitude larger than recent estimates of $\sim$ 10$^3$ g s$^{-1}$, based on Jeans escape rates at exobase temperatures close to $\te$ (Sasselov 2003). 

Concerning the case of HD209458b, we cannot find an evaporating sequence reproducing
its properties. As for non-evaporating models (see B03), the predicted radius
for the observed mass and inferred age of the system is about 25 \% smaller than the observed value.
Interestingly enough, we find that a planet at $a$ =0.046 AU with initial mass $\sim$ 1.1-1.2 $\mjup$ reaches the critical regime $t_{\rm \dot M} / t_{\rm KH} < 0.1$ precisely at the age of HD209458b ($\log \, t = 9.6-9.85$) (see Fig. \ref{fig2}). 
Since our analysis suggests a violent reaction of
a planet when reaching this regime, with rapid expansion of
the outer layers (see Fig. 1),
we may wonder whether HD209458b is not in such a regime. Although the probability
to see a transit planet in this rapid phase is very small, the discovery of other similar systems with unexplained large radii would suggest further attention to this scenario.

\section{Conclusion}

The present calculations, including a consistent treatment of irradiation and evaporation due to XUV/Lyman-$\alpha$ irradiation during
the evolution of irradiated planets, suggest the existence of a critical mass $m_{\rm crit}$,
which varies with orbital separations, below
which the evaporation timescale becomes significantly shorter than the thermal timescale
of the planet. 
%This critical mass depends on the orbital distance. 
Based on the present evaporation rates for solar conditions, we get $m_{\rm crit} \sim$ 1.5 $\mjup$ at  $a$ = 0.046 AU and $m_{\rm crit} \sim 2.7 \mjup$ at $a$ = 0.023 AU. 
For objects with mass below this critical mass, we find that:
(i) the planet will evaporate entirely (except possibly for the central rocky core) within about 5 Gyr;
(ii) its lifetime is shortened by a factor $\sim$ 2 compared with the time
predicted for complete evaporation but omitting the effect on the evolution;
(iii) its outer layers expand rapidly and its radius eventually increases at some time,
(iv) the planet undergoes a phase of rapid mass loss which could expel part or all of its remaining hydrogen-rich envelope in a very short timescale.
Planets with initial masses $m_{\rm i}>m_{\rm crit}$ survive to evaporation on a lifetime longer than 5 Gyr, and evolution is similar to the 
case of a non-evaporating irradiated planet.
The values of $m_{\rm crit}$
depend on the evaporation rates which are still very uncertain and rely on a rather primitive
model for such a complex process. Moreover, applying
 evaporation rates determined at the exosphere to the photospheric surface of the planet, the relevant boundary for evolution, is an extreme simplification. %Finally, non-thermal effects as discussed in L03 may modify the total evaporation rate.
%may result in a weak magnetic shielding against stellar wind, thus reducing evaporation.
%Although this model and the present choice of parameters may  overestimate the thermal evaporation effects that 
%can be expected from XUV/Lyman-$\alpha$ irradiation, non-thermal effects as discussed in L03
%may instead results in a weak magnetic shielding against stellar wind, thus reducing evaporation.
%may also play an important role, in particular in the case of tidally-locked hot jupiters. Their atmospheric expansion results in a weak magnetic shielding against stellar wind (Griessmeier et al. 2003). 
%It is thus difficult to determine what the exact evaporation rates will be, as the outcome of these complex, intricated non-linear processes.
%Realisitic evaporation rates may thus differ appreciably from present estimates, even if the latter provide a good agreement
%with observations of HD209458b. 
Uncertainties on the mass loss rates, however, do not affect the qualitative
existence of a planet critical mass, for a given orbital distance, below which a planet eventually will react violently to evaporation and will expand again, possibly until complete evaporation.
Such a behavior bears important consequences on the lifetime
of close-in planets and on our understanding of their mass-orbital period distribution. 
%It may also explain the surprisingly large radius of HD209458b. 
Our results thus provide an excellent motivation to pursue efforts to understand evaporation effects and to explore further the influence of the high energy part of the parent star
incident flux.


\begin{thebibliography}{}

\bibitem[]{} Armitage, P., Cathie, J. \& Palla, F., 2003, MNRAS, 342, 1139
\bibitem[]{} Baraffe, I., Chabrier G., Barman, T., Allard F., Hauschildt P
.H., 2003, A\&A, 402, 701 (B03)
\bibitem[2001] {barman} Barman, T., Hauschildt, P.H., Allard, F. 2001, \apj, 
556, 885
%\bibitem[]{} Bailey, S. M.,  Woods, T. N.,  Canfield, L. R., Korde, R.,  Barth, C. A., Solomon, S. C., Rottman,G. J., 1999, Sol. Phys, 186, 243
%\bibitem[]{} Burrows, A., Sudarsky, D., Hubbard, W.B. 2003, \apj, 594, 545 
%\bibitem[]{} Chabrier, G., Baraffe, I. 1997, \aap, 327, 1039
\bibitem[]{} Chabrier G., Barman, T., Baraffe, I.,  Allard F., Hauschildt, P., 2004, \apjl, 603, L53
\bibitem[]{} Charbonneau, D., Brown, T., Latham, D., Mayor, M. 2000,
\apj, 529, L45
%\bibitem[2002]{cody} Cody, A.M., Sasselov, D.D. 2002, \apj, 569, 45
%\bibitem[]{} Floyd,L., Prinz, D., Crane, P., Herring, L., 
%Brueckner, G., 1999, Adv. Space Res., 24-2, 225
%\bibitem[2002]{guillot} Guillot, T., Showman, A.P. 2002, \aap, 385, 156
\bibitem[1996]{guillot96} Guillot, T., Burrows, A., Hubbard, W.B., Lunine, J.I., Saumon, D. 1996, \apj, 459L
%\bibitem[]{} Griessmeier, J.-M., Motschmann, U., Stadelmann, A., Penz, T., Lammer, H., Selsis, F., Ribas, I., Guinan, E. F., Biernat, H. K., Weiss, W. W., 2003, submitted to A\&A
\bibitem[]{} Konacki, M., Torres, G., Jha, S., Sasselov, D., 2003, Nature, 421, 507
\bibitem[]{} Lammer, H., Selsis, F., Ribas, I., Guinan, E., Bauer, S.J., Weiss, W.
2003, ApJL, 598, L121 (L03)
%\bibitem[]{} Lean, J., Warren,ÊH., Mariska,ÊJ., Bishop,ÊJ., 2003,
%J. Geophys. Res., 108, A2, 1059
%\bibitem[]{} Mazeh, T., Naef, D., Torres, G. et al. 2000, ApJ, 532, L55
\bibitem[]{} Ritter H. 1996, in
Evolutionary Processes in Binary Stars, Kluwer,
Dordrecht, NATO ASI series, Vol. 477, p.~223
\bibitem[]{} Torres, G., Konacki, M., Sasselov, D., Jha, S. 2003, APJL, submitted 
(astro-ph/0310114)
\bibitem[]{} Sasselov, D. 2003, APJ, 596, 1327
%\bibitem[]{} Stehle R., Ritter H., Kolb U. 1996, MNRAS, 279, 581
\bibitem[]{} Strom, S., Edwards, S., Skrutskie, M.,1993, in Protostars and planets III, p. 837
\bibitem[]{} Vidal-Madjar, A., Lecavalier des Etangs, A., D\'esert, J-M., Ballester, G.,
Ferlet, R., H\'ebrand, G., Mayor, M. 2003, Nature, 422, 143
\bibitem[]{} Watson, A., Donahue, T., Walker, J., 1981,
Icarus, 48, 150
\bibitem[]{} Woods, T.N., Rottman, G.J. 2002, 
 in Atmospheres in the Solar System: 
Comparative Aeronomy, Geophysical Monograpg 130, AGU, p.221
%\bibitem[]{} Woods, T.,  Eparvier, F., Solomon, S., Woodraska, D., \& Bailey, S., 2002, in 4th Thermospheric/Ionospheric Geospheric Research (TIGER) Symposium
%Solar-EUV-Experiment on TIMED mission
\end{thebibliography}
\end{document}